 \documentclass[manuscript]{acmart}
 
\AtBeginDocument{%
  \providecommand\BibTeX{{%
    \normalfont B\kern-0.5em{\scshape i\kern-0.25em b}\kern-0.8em\TeX}}}

\setcopyright{acmcopyright}
\copyrightyear{2018}
\acmYear{2018}
\acmDOI{XXXXXXX.XXXXXXX}

\acmConference[ACM XXX XXX]{Make sure to enter the correct
  conference title from your rights confirmation emai}{June 03--05,
  2018}{Woodstock, NY}
\acmBooktitle{Woodstock '18: ACM Symposium on Neural Gaze Detection,
 June 03--05, 2018, Woodstock, NY} 
\acmPrice{15.00}
\acmISBN{978-1-4503-XXXX-X/18/06}

\newcommand{\yashar}[1]{\textcolor{magenta}{{\bf [Yashar: }{\em #1}{\bf ]}}}

\newcommand{\laPFR}{\texttt{LaPFR\:}}
\newcommand{\maPFR}{\texttt{MaPFR\:}}
\newcommand{\haPFR}{\texttt{HaPFR\:}}
\newcommand{\tPFR}{\texttt{tPFR\:}}

\newcommand{\dquotes}[1]{``#1''}
\newcommand{\squotes}[1]{`#1'}
\usepackage{xcolor, colortbl}
\usepackage{comment}
\definecolor{green(pigment)}{rgb}{0.0, 0.65, 0.31}
\definecolor{emerald}{rgb}{0.31, 0.78, 0.47}
\definecolor{cadmiumgreen}{rgb}{0.0, 0.42, 0.24}
\definecolor{blue(pigment)}{rgb}{0.2, 0.2, 0.6}

\usepackage{algpseudocode}
\usepackage{algorithm}
\usepackage{multirow}
\usepackage{multicol}
\usepackage{makecell}

\begin{document}

\title[Fairness for All:]{Fairness for All: Investigating Harms to Within-Group Individuals in Producer Fairness
Re-ranking Optimization - A Reproducibility Study}


\author{Giovanni Pellegrini}
\email{g.pellegrini4@studenti.poliba.it}

\author{Vittorio Maria Faraco}
\email{v.faraco@studenti.poliba.it}

\author{Yashar Deldjoo}
\email{yashar.deldjoo@poliba.it}
\affiliation{%
  \institution{Politecnico di Bari}
  \city{Bari}
  \country{Italy}
}

\renewcommand{\shortauthors}{G. Pellegrini et al.}

\begin{abstract}

Recommender systems are widely used to provide personalized recommendations to users. Recent research has shown that recommender systems may be subject to different types of biases, such as popularity bias, leading to an uneven distribution of recommendation exposure among producer groups. To mitigate this, producer-centered fairness re-ranking (PFR) approaches have been proposed to ensure equitable recommendation utility across groups. However, these approaches overlook the harm they may cause to within-group individuals associated with colder items, which are items with few or no interactions.

This study reproduces previous PFR approaches and shows that they significantly harm colder items, leading to a fairness gap for these items in both advantaged and disadvantaged groups. Surprisingly, the unfair base recommendation models were providing greater exposure opportunities to these individual cold items, even though at the group level, they appeared to be unfair. To address this issue, the study proposes an amendment to the PFR approach that regulates the number of colder items recommended by the system. This modification achieves a balance between accuracy and producer fairness while optimizing the selection of colder items within each group, thereby preventing or reducing harm to within-group individuals and augmenting the novelty of all recommended items. The proposed method is able to register an increase in sub-group fairness (SGF) from 0.3104 to 0.3782, 0.6156, and 0.9442 while also improving group-level fairness (GF) (112\% and 37\% with respect to base models and traditional PFR). Moreover, the proposed method achieves these improvements with minimal or no reduction in accuracy (or even an increase sometimes). 

We evaluate the proposed method on various recommendation datasets and demonstrate promising results independent of the underlying model or datasets. Our reproducibility study highlights the importance of considering within-group individuals in fairness-improving approaches and proposes a potential solution to address the issue of harm to disadvantaged individuals. We believe that our proposed method can contribute to ongoing efforts to make recommender systems more inclusive and fair to all users.

\end{abstract}

\received{February 2023}
\received[revised]{August 2023}
\received[accepted]{August 2023}


\maketitle

\section{Introduction and context}

Recommender systems are widely used in task-sensitive and business-competitive domains, such as employment, health, e-commerce, and social media. As a result, fairness has gained increasing importance in recent years. The field of recommender systems typically examines fairness from two perspectives: consumer-provider (stakeholder) and group-individual (granularity of groups) \cite{deldjoofairness,wang2023survey}. A recent survey \cite{deldjoofairness}, which reviews fairness studies in recommender systems up to 2023, states that group fairness has emerged as the predominant focus of research in RecSys, accounting for 67\% of all studies. Group fairness refers to ensuring fairness across a specific demographic, while individual fairness aims to treat each person as an individual, regardless of group membership. The group fairness approach is further categorized into Consumer Fairness (CF) and Producer Fairness (PF), with almost 50\% of all studies in fairness research dedicated to these two subcategories.

In group fairness setting, a key assumption is that providing equitable recommendation utility at the \textbf{group level} is sufficient to deem the method fair. For PF hence, this implies that both privileged and underprivileged provider groups should receive comparable exposure according to a predetermined target distribution (e.g., equal or proportional to catalog size). In this scenario, it can be  even acceptable to compromise some recommendation accuracy to maintain the equity of recommendation exposure at the group level.

Various fairness enhancement techniques have been developed to achieve this goal, including pre-processing, in-processing, and post-processing methods~\cite{pitoura2022fairness, mehrabi2021survey, shrestha2019fairness}. Our study focuses on post-processing fairness ranking techniques that are adaptable to different recommendation algorithms and contexts without being tied to the core recommendation algorithm, also referred to as the \dquotes{base ranking model}. These techniques have gained significant attention due to their ability to transform black-box ranking methods into fair rankings and since they do not require re-training if fairness or protected groups change. This feature is particularly useful when group and fairness definitions are dynamic and re-training a model is costly. We review here several applications of fairness ranking approaches in recommender systems research.~\citet{ferraro2021break} propose a re-ranking algorithm that prioritizes user-oriented fairness and explicitly addresses gender and music bias. They evaluated the algorithm's effectiveness on a limited range of base ranking models, which included collaborative filtering approaches and well-known baselines such as MostPop.~\citet{yalcin2021investigating} address popularity bias in \textit{group recommendations} by adapting a re-ranking approach and proposing two strategies that incorporate popularity and group ratings. Similarly,~\citet{li2021user} present a user-oriented fairness re-ranking (UFR) method to address the unfairness problem in recommendations by adding constraints to evaluation objectives in the optimization algorithm.~\citet{rahmani2022experiments} extend these studies by examining UFR across different group fairness definitions and settings based on attributes and domains. In a similar work,~\citet{naghiaei2022cpfair} also explore the applicability of fairness re-ranking in various scenarios, such as provider and joint consumer-provider fairness re-ranking, using a large number of datasets. These research studies highlight the versatility of fairness re-ranking optimization in addressing algorithmic biases and promoting fairness in various recommendation and fairness settings. Our study aims to investigate the \textbf{harms and consequences} of \textit{producer-centered fairness re-ranking (PFR)} on individuals within a group, with a particular emphasis on the exposure of less popular items in sub-groups. The main objective is to determine if the \tPFR method successfully achieves the intended producer's goal of enhancing the visibility of these items.

We suggest a modification to the \tPFR algorithm for improving both group-level and sub-group-level fairness, which we evaluate using the mean novelty of less popular items in sub-groups as a metric to maximize. The revised approach is referred to as \texttt{\textbf{aPFR}} and is introduced in various versions in this work.\footnote{While the focus of this work is on provider fairness, the findings and insights obtained may have wider implications for other fairness scenarios, including consumer fairness and two-sided markets.}



\begin{figure*}[t]
    \centering
    \includegraphics[width=\textwidth]{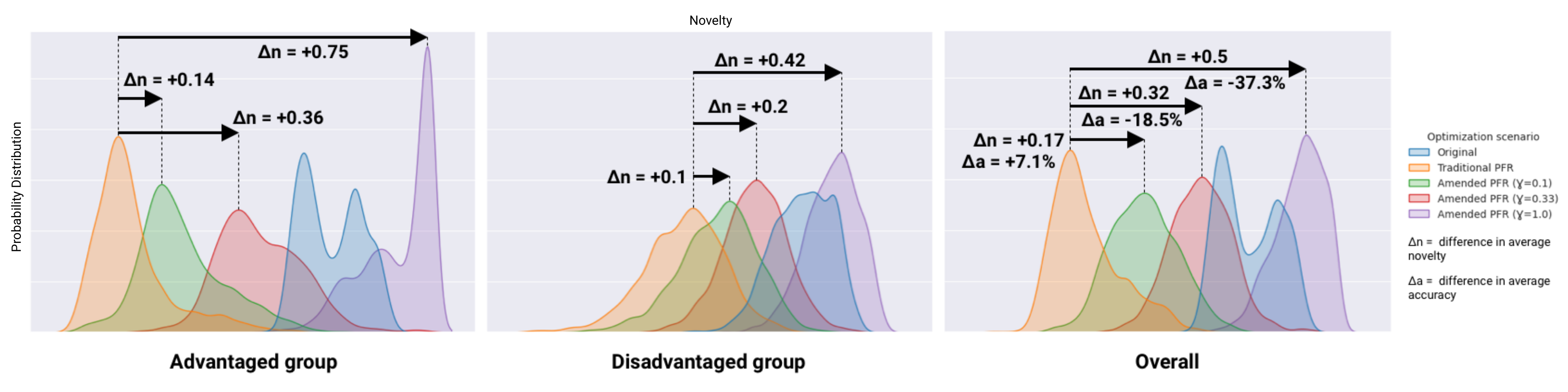}
    \caption{Comparison of Concentration Bias in Novelty Distribution of User Recommendation Lists between Traditional provider fairness re-ranking (\tPFR) and Amended PFR: Analysis on \texttt{Amazon Luxury Beauty} using the MultiVAE model. (\textbf{Left}) Within the advantaged group, (\textbf{Middle}) within the dis-advantaged group, (\textbf{Right}) overall in recommendation lists. $\Delta_n$ measures the improvement in the novelty of the recommended item, while $\Delta_a$ quantifies the cost/gains in terms of the recommendation accuracy.}
    \label{fig:con_bias}
\end{figure*}

\begin{table*}[t]
\centering
\caption{Cost-Benefit Analysis in Fig. \ref{fig:con_bias}'s optimization scenarios for Group Fairness (PFR goal), Accuracy and Sub-group Fairness (PFR costs), on the \texttt{Amazon Luxury Beauty} dataset, with MultiVAE and NeuMF as baseline models. Note that novelty is directly related to the fairness, which we analyze it at sub-group level referred in this work as sub-group cold-item fairness .}
\label{tbl:con_bias}
\small
\scalebox{0.875}{\begin{tabular}{l|c|llllll|lllll}
\toprule
\multirow{2}{*}{Metric} & \multirow{2}{*}{Role} & \multicolumn{5}{c}{\textbf{MultiVAE} \textit{(plot in Figure 1)}} && \multicolumn{5}{c}{\textbf{NeuMF}} \\
\cmidrule{3-7}\cmidrule{9-13}
                        &                       & \cellcolor{blue!20} Base & \cellcolor{orange!20} \tPFR & \cellcolor{green!20} \laPFR& \cellcolor{red!20} \maPFR & \cellcolor{violet!20} \haPFR && \cellcolor{blue!20} Base & \cellcolor{orange!20} \tPFR & \cellcolor{green!20} \laPFR & \cellcolor{red!20} \maPFR & \cellcolor{violet!20} \haPFR \\
\midrule
\multicolumn{13}{c}{\textbf{}} \\ \hline
\textbf{Group Fairness} $\downarrow$  & Goal & 46.18 & 22.56  & \textbf{6.42} & \textbf{1.1}  & \textbf{2.4}  && 45.51 & 2.26  & 2.26  & 2.26  & 2.26  \\
\textbf{Accuracy} $\uparrow$ & Cost 1 & 0.00934 & 0.00869  & \textbf{0.00931} & 0.00708  & 0.00545  && 0.01152 & 0.01017  & \textbf{0.00951}  & \textbf{0.00916}  & 0.00789  \\
\textbf{Novelty Adv. Group} $\uparrow$ & \textit{ - } & 0.72 & 0.16  & \textbf{0.3}  & \textbf{0.52}  & \textbf{0.91} && 0.74 & 0.15  & \textbf{0.24} & \textbf{0.32}  & \textbf{0.45}  \\
\textbf{Novelty Disadv. Group} $\uparrow$ & \textit{ - } & 0.76 & 0.43 & \textbf{0.53} & \textbf{0.63} & \textbf{0.85} && 0.79 & 0.55 & \textbf{0.55} & \textbf{0.57} & \textbf{0.61} \\
\textbf{Avg. Sub-Group Nov.} $\uparrow$ & Cost 2 & 0.74 & 0.29 & \textbf{0.42} & \textbf{0.58} & \textbf{0.88} && 0.77 & 0.35 & \textbf{0.4} & \textbf{0.45} & \textbf{0.53} \\
\bottomrule
\end{tabular}}
\begin{flushleft} Our objective in this work is to maximize the visibility of cold items within sub-groups, and we use the metric {\small \textbf{Avg. Sub-Group Nov.}} to measure this. This aligns with the provider's goal of increasing exposure to less popular items while maintaining accuracy (cf. Section \ref{subsec:fairness-metrics}). We use the terms \dquotes{sub-group cold-item fairness} and \dquotes{sub-group fairness} interchangeably in this work to refer to the same notion of fairness.
\end{flushleft}
\vspace{0.325em}
\hrule
\end{table*}

\vspace{0.5em}\noindent \textbf{\textit{The Fairness Harms Resulting from \tPFR.}} It is crucial to re-evaluate the underlying premise of group fairness and analyze the \textbf{implications} of the group fairness provided by PFR. While it is natural to assume that all these fairness interventions are aimed at promoting fairness in recommendations to providers, we must scrutinize this assumption more closely and determine if we have been successful in promoting fairness in society. To accomplish this, we will examine a specific example, as demonstrated in Figure \ref{fig:con_bias} and Table \ref{tbl:con_bias}. 

Figure \ref{fig:con_bias} demonstrates the distribution of item novelty in the recommendation lists produced by the MultiVAE model on the \texttt{Amazon Luxury Beauty} dataset used in our experiments. The figure comprises five plots, each representing the novelty distribution across users. Table \ref{tbl:con_bias} presents the numerical values for the evaluation metrics, which are calculated for two different CF models: MultiVAE and NeuMF. It is worth noting that the traditional PFR, as explored in \cite{naghiaei2022cpfair} and originally derived from the work of \citet{li2021user}, aims to enhance Group Fairness (Goal) while minimizing the impact on Accuracy (Cost 1). The current work at hand introduces the concept of \dquotes{sub-group cold-item} fairness as an additional cost of the PFR intervention (Cost 2) to be measured.

Table \ref{tbl:con_bias} shows that \tPFR successfully achieves a significant reduction in group-level fairness from 46.18 to 22.56, which is more than a 51\% reduction, at the cost of a moderate sacrifice in accuracy from 0.00934 to 0.00869 (or 7\%). However, the results also reveal that \tPFR causes a substantial reduction in the fairness of items within each sub-group, from 0.74 to 0.29, which is a 61\% reduction.\footnote{It is important to note that the measurement of sub-group fairness is based on a logarithmic scale, which means a twofold difference corresponds to a tenfold difference on the normal scale.} In simpler terms, the outcome of \tPFR is as follows: it achieves group-level fairness (Goal 1, +51\%), with a trade-off in accuracy (Cost 1, -7\%). However, it harms the exposure of cold-items within each sub-group, resulting in a reduction of sub-group fairness (Cost 2, -61\%). Therefore, even though \tPFR appears fair at the group-level and increases the exposure of disadvantaged item group (less popular items), it harms the exposure of cold items within each sub-group, which contradicts the goal of provider fairness (cf. Section \ref{subsec:fairness-metrics}).

To address this issue, we propose an amendment to PFR, called \texttt{aPFR}, which introduces a multiplicative novelty term into the re-ranker regularization function to control the novelty of recommended items. These plots are arranged from left to right in the following order:

\begin{itemize}
    \item  {\color{blue} \textbf{Blue curve.}}
    The baseline ranking model (before fairness);
    \item {\color{orange} \textbf{Orange curve.}} The \tPFR algorithm, which was originally derived from a study by \citet{li2021user}, and explored in \cite{naghiaei2022cpfair};
    \item {\color{green(pigment)} \textbf{Green curve.}}  It represents the proposed amendment with light emphasis (\laPFR) on long-tail items within each sub-group ($\gamma = 0.1$);
    \item {\color{red} \textbf{Red curve.}} It represents the proposed amendment with medium emphasis (\maPFR) on long-tail items within each sub-group ($\gamma = 0.33$);
    \item {\color{violet} \textbf{Violet curve.}} It represents the proposed amendment with heavy emphasis (\haPFR) on long-tail items within each sub-group ($\gamma = 1$).
\end{itemize}
The parameter $\gamma$ is used in the re-ranker to regulate cold-item exposure, and our evaluation shows that our method achieves higher group-level fairness and more exposure for less popular items in sub-groups compared to \tPFR, by simultaneously regularizing group-level fairness and cold-item exposure.


The Light-, Medium-, and Heavy-Amended PFR (\laPFR, \maPFR, and \haPFR) - shown in green, red, and violet curves, respectively - promote sub-group fairness, increasing it from 0.27 to 0.42, 0.58, and 0.88, respectively, compared to \tPFR. Additionally, they improve or maintain both group-level fairness (from 22.56 to 6.42, 1.1, and 2.4) and accuracy (from 0.00869 to 0.00931, 0.00708, and 0.00545), as shown in Table \ref{tbl:con_bias}. For example, the \laPFR approach not only reduces fairness with respect to \tPFR, from 22.56 to 6.42 (71.61 \% reduction) but also increases accuracy, from 0.00869 to 0.0931 (7.14 \%). This increase in accuracy can be attributed to the larger search space in \laPFR (or in general aPFR), which is able to retrieve good items from long-tail distributions in each sub-group. As we increase the power of novelty, we are able to enhance sub-group fairness and group fairness, however, this comes at the cost of overall system accuracy. By exploring the aPFR approaches, we are able to identify the \dquotes{sweet spots} where we can maintain or improve accuracy while enhancing fairness at different levels.

\vspace{0.5em}\noindent \textbf{\textit{Summary.}}
According to the studies and analyses conducted, the pursuit of group-level fairness through \tPFR can result in reduced exposure of less popular items (i.e., less fairness), which contradicts the intended purpose of these methods, even if the algorithm considers it fair. We offer an explanation for this unfairness at the sub-group level, which we believe is caused by the fixed threshold of binary grouping, resulting in greater harm to cold items within sub-groups, possibly due to user preference for short-head items in sub-groups. Our proposed solution effectively tackles this issue by incorporating a multiplicative novelty term into the re-ranker cost function. Through extensive empirical evaluation, it has been demonstrated that the combination of the novelty term and group-level fairness term yield enhancements in \textbf{accuracy}, \textbf{group fairness}, and the proposed \textbf{sub-group fairness}. Moreover, depending on the task and interest of the stakeholder (e.g., provider), the proposed solution can further enhance fairness at both levels by prioritizing cold items in \maPFR and \haPFR, at the expense of a mild reduction in accuracy.

\vspace{0.5em} \noindent \textbf{\textit{Contributions.}}
In this context, this research aims to reproduce previous \tPFR models, paying specific attention to the harm they do to within-group individuals associated with colder items, in particular

\begin{itemize}
    \item We conduct a thorough analysis of the commonly used provider fairness re-ranking (PFR) approach in prior research and draw attention to its negative impacts on exposure of cold items when scrutinized at a more granular group level;
    \item We propose an amendment to the classic \tPFR approach by introducing a \textit{novelty multiplier} term to the cost function to control the degree of novelty in the recommended items. Our approach balances group-level fairness and sub-group fairness by fine-tuning the novelty multiplier and its interaction with fairness regularization. This enables us to promote the visibility of cold items in sub-groups while maintaining high accuracy levels;
     \item We conducted an ablation study where we tested various combinations of regularization parameters for both the fairness parameters (capturing group-level fairness) and the cold-start term (capturing sub-group fairness). The objective was to improve the generalizability of our research and investigate the interaction between the amended \tPFR method and the ablation parameters.
    
    


    \item We conduct extensive experiments by applying re-rankers on top of various competitive baseline collaborative filtering (CF) recommendation approaches. Specifically, we explore five domains, namely movie (\texttt{MovieLens100K}), luxury-wellness (\texttt{Amazon Luxury Beauty}), e-commerce (\texttt{Amazon Prime Pantry}), POI (\texttt{Foursquare}), and music (\texttt{LastFM}), which have different feedback types, including explicit and implicit. We evaluate our the proposed \texttt{aPFR} amendments on five diverse baseline recommendation models: BPR, NeuMF, MultiVAE, LightGCN, and NGCF, giving us a total combination of 5 datasets $\times$ 5 models $=$ 25 CF simulations. 

\end{itemize}

In the following, we investigate the reproducibility technique (cf. Section \ref{sec:rep}), then we replicate prior experiments with the proposed amendment (cf. Section \ref{sec:exp}), we highlight the results and findings of our research (cf. Section \ref{sec:findings}), and we draw consequent conclusions that pave the way for more fair future work (cf. Section \ref{sec:conc}). 

\section{Reproducibility technique}
\label{sec:rep}
In this section, we define the PFR implementation we use and reproduce, we present the datasets we run our experiments on, the base ranking models we employ, the fairness definitions our work is based upon, and its evaluation methods and metrics.

\subsection{Producer Fairness Re-ranking (PFR)}

\subsubsection{Background}
The optimization-based re-ranking approach used in this study, known as \tPFR, has been previously investigated in recommendation system research, as discussed in \cite{rahmani2022experiments,li2021user,naghiaei2022cpfair}. The primary differences between these strategies stem from their focus on different stakeholders and their use of constrained or unconstrained optimization strategies. Initially, \citet{li2021user} considered using an unconstrained optimization-based re-ranking approach for consumer fairness in CF settings. However, subsequent research \cite{rahmani2022experiments,naghiaei2022cpfair} focused on consumer or CP-fairness by utilizing a constrained approach. In our study, we built upon the constrained version of the approach and focused on \textit{provider fairness}. We chose this focus for two main reasons: \textit{firstly}, it allowed for a deeper understanding of the method and what PFR achieves, particularly given that PFR operates at a user level rather than a CP level, and \textit{secondly}, encouraging the promotion of cold-items has notable commercial advantages, which is in line with the motivation behind provider fairness. (cf. Section \ref{subsec:fairness-metrics}). 

\subsubsection{Formal description of the proposed amendment to PFR}
To provide context for our proposed amendment, we present a brief overview of the PFR method. PFR uses the top-N recommendation list and relevance scores from the unfair base ranker and applies an optimization-based re-ranking algorithm. The objective is to maximize the total relevance scores while minimizing the deviation from producer fairness ($\mathbf{GF}$). This approach has been explored in previous studies on recommendation systems, such as those discussed in \cite{rahmani2022experiments,li2021user,naghiaei2022cpfair}.

The re-ranking optimization objective can be formalized as follows, with the decision vector $X$ selecting the items to be included in the re-ranked list:

\begin{small}
\begin{equation}
\begin{aligned}
\max_{X_{i}} \quad & \sum_{i=1}^{N}{S_{i} \cdot \textbf{N}_{i}^{\gamma} \cdot X_{i}} - \lambda \cdot \textbf{GF}(X,\mathcal{I})\\
\textrm{s.t.} \quad & \sum_{i=1}^{N}{X_{i}} = K, X_{i} \in{\{0,1\}}\\
\end{aligned}
\label{eq:optimization}
\end{equation}
\end{small}

The optimization problem aims to maximize total preference scores $\mathbf{S_i}$ and minimize deviations from fairness by recommending a specific number $K$ from the top-$N$ recommendation list of items to each user that minimizes $\mathbf{GF}$. Similar to \cite{naghiaei2022cpfair}, $\mathbf{GF}$ here computes the difference between the expected recommendation utility of items in the advantaged group and that of items in the disadvantaged group. In addition, we extend the use of $\mathbf{GF}$ by introducing a weighted version that compares the deviation from parity exposure to a target distribution. This enables us to simultaneously consider the fairness concerns of multiple producer groups and evaluate the performance of the recommender system against a pre-defined fairness target (cf. Section \ref{subsec:fairness-metrics}).

The term $\textbf{N}$ introduces the novelty dimension in the re-ranker objective by influencing the selection of items that have a higher novelty score, enabling the method to tackle sub-group fairness jointly with the other goals. The hyperparameters $\lambda$ and $\gamma$ determine the emphasis placed on group fairness deviation and sub-group cold-item fairness respectively, with a value of 0 aligning with the baseline recommendation list. The optimization problem mentioned is a mixed-integer linear programming (MILP) problem and is known to be NP-hard. However, commercial optimization solvers such as Gurobi can be used to solve it. The MILP problem can be converted to an instance of the Knapsack problem, where the objective is to select items for each user to maximize the overall score, with the assumption that each item has a unit weight and the total weight is limited by the fixed list size.

\subsubsection{Implementation.}
The implementation of \tPFR \cite{naghiaei2022cpfair} is publicly available on Google Colab via their Github repository.\footnote{\href{https://github.com/rahmanidashti/CPFairRecSys}{https://github.com/rahmanidashti/CPFairRecSys}} It utilizes Cornac,\footnote{\href{https://cornac.preferred.ai}{https://cornac.preferred.ai}} a recommendation framework, and the optimization framework MIP\footnote{\href{https://www.python-mip.com}{https://www.python-mip.com}}, which is based on the commercial optimization solver Gurobi.\footnote{\href{https://www.gurobi.com}{https://www.gurobi.com}} The code includes implementation steps for training baseline models, testing pipelines, and fairness-aware re-ranking and evaluation modules. However, due to Colab's recent discontinuation of support for TF1, their code cannot be executed anymore, which seriously hinders its reproducibility aspect. For instance, running neural models like NeuMF is no longer possible. 
To address this issue, we employ RecBole,\footnote{\href{https://recbole.io}{https://recbole.io}} a highly flexible recommendation framework built on PyTorch, in our study. We are making our implementation publicly accessible through a dedicated anonymized repository, which is discussed in more detail in Section \ref{subsec:code}.

\subsection{Fairness Definitions and Metrics} \label{subsec:fairness-metrics}
The focus of this work is on provider fairness and explores fairness within this context. Before delving into the analysis, we revisit the primary fairness goal in provider fairness and then examine the level of fairness within this framework.

\vspace{1em}\colorbox{gray!10}{
\begin{minipage}{0.94\textwidth}
\noindent \textbf{Goal. \dquotes{Provider's Objective of Fairness in Recommender Systems}}

From a provider's perspective, the ideal goal is to enhance the exposure of cold or less popular items offered by the RS while simultaneously ensuring an acceptable level of recommendation quality, which means maintaining recommendation accuracy with little or no reduction.
\end{minipage}} \\ 

We investigate fairness at two \textit{hierarchical grouping levels} that have received limited attention in prior works to achieve the above objective. Specifically, we aim to examine the goal of provider fairness for the exposure of cold items at different levels of granularity, namely, the group and sub-group levels. These levels are elaborated below:

\noindent \begin{definition}[\textbf{Group-level provider fairness}]
\textit{In the context of group-level fairness, a recommender system is deemed fair towards producers if it offers equitable recommendation utility or exposure to both privileged and underprivileged groups, as determined by a \dquotes{target representation}.} 
\end{definition}

\noindent Note that target representation refers to the ideal proportion or distribution of exposure of different groups in a recommendation system, as discussed in \cite{kirnap2021estimation}. This paper considers two target representations for group fairness in recommender systems: (i) parity, and (ii) proportionality to corpus presence. Parity aims for equal resource allocation for each group, while proportionality targets allocation proportional to the number of items in the corpus belonging to a given group. We use a popularity-based segmentation approach, categorizing items as either short-head or popular items (top 20\%) or long-tail or unpopular items (bottom 80\%). The re-ranking objective integrates group-level fairness by enhancing the exposure of items from both groups based on a target representation (fair distribution) specified by the system designer. The target representations used in this work are denoted as $GF_{eq}$ and \textbf{$GF_{prop}$}, which correspond to $p_f$ values of $[0.5, 0.5]$ and $[0.2, 0.8]$, respectively.

\begin{definition}[\textbf{Sub-group cold item fairness}]
\textit{Sub-group fairness, as defined in the context of provider fairness, refers to the level of fairness at an individual level within each sub-group. In our approach, we specifically focus on "cold items", where we assume that increased exposure to these items corresponds to greater sub-group fairness. This definition aligns with the objective of provider fairness, which aims to enhance the visibility of less popular items.}
\end{definition}

We group each recommendation list into two sub-groups, $G = [G_A, G_B]$, where $G_A$ and $G_B$ represent popular and non-popular item groups. Then we compute a novelty score using the following formula: $ N(i | i\in\mathcal{C}) = \sum_{i\in\mathcal{C}}-log_2(p_i) $, where $p_i$ is the popularity score of item $i$ in the original catalog $C$. We denote the novelty score for sub-group $G_A$ and $G_B$ as $n_A$ and $n_B$, respectively. Then, we calculate the novelty scores for both sub-groups and then compute the average of them to obtain the SGF score (short for Sub-group cold-item fairness), represented by the formula $SGF = (nA + nB)/2$. A higher SGF value indicates that both sub-groups provide more exposure to colder items within their respective groups, thereby achieving better SGF.

\subsection{Code and Datasets}
\label{subsec:code}
The entire pipeline code, which includes training and inference of the base ranking models and the fairness-aware re-ranking stage, along with pre-processed data used to generate the results in this paper, is released. Furthermore, we offer ready-to-run Jupyter Notebooks that have been tested on Google Colab to obtain the results and plots mentioned in the paper. All the relevant materials can be found on the anonymized GitHub repository.\footnote{\href{https://anonymous.4open.science/r/aPFR-SGF-2C56/README.md}{https://anonymous.4open.science/r/aPFR-SGF-2C56/README.md}}

\section{Replicating Prior Experiments with the Proposed Amendment.}
\label{sec:exp}

This section outlines experiments conducted to assess various reproducibility aspects, including the obtained results and our observations.



\subsection{Setting.}
Below are all the details concerning the experimental setup.
\subsubsection{Datasets}
We employ five publicly available datasets coming from different domains, that are movie, luxury-wellness, e-commerce, POI, and music, and contain different feedback types (i.e., explicit and implicit). The datasets are \texttt{MovieLens-100K}, \texttt{Amazon Luxury Beauty}, \texttt{Amazon Prime Pantry}, \texttt{Foursquare}, and \texttt{LastFM}. Before using them, we apply \textit{k}-core data filtering with $k = 10$ to ensure they contain sufficient number of ratings per user ($\frac{R}{U}$) and ratings per item ($\frac{R}{I}$), leading to a higher density ($\frac{R}{U \times I}$) and ultimately a more manageable overall size to run our experiments in a dynamic setting with enough feedback. The datasets statistics are described in detail in Table \ref{tbl:datasets}. 

\begin{table}
\caption{Statistics of the final datasets used in this work after $k$-core pre-processing.}
\label{tbl:datasets}
\centering
\begin{tabular}{lccccccc}
\toprule
\textbf{Dataset} & Users & Items & Ratings & $\frac{R}{U}$ & $\frac{R}{I}$ & Density & Item Gini \\
\midrule
\textbf{MovieLens100K}  & 944 & 1,683 & 100,000 & 106.04 & 59.45 & 0.063\% & 0.629 \\
\textbf{Amazon Prime Pantry}  & 6,049 & 4,367 & 78,186 & 12.91 & 17.9 & 0.003\% & 0.279 \\
\textbf{Amazon Luxury Beauty} & 2,719 & 1,028 & 18,466 & 6.8 & 18 & 0.006\% & 0.299 \\
\textbf{Foursquare} & 1,083 & 5,135 & 147,938 & 136 & 29 & 0.027\% & 0.422 \\
\textbf{LastFM} & 1,867 & 1,529 & 62,795 & 33.73 & 41.2 & 0.022\% & 0.529 \\
\bottomrule
\end{tabular}
\end{table}

\subsubsection{Evaluation Method.}
Our settings involve a train-validation-test split for the data with ratios 80\%, 10\%, and 10\%, respectively. In contrast to \cite{li2021user}, we evaluate our approach under the assumption that the relevance scores are not accessible to the post-processing algorithm, and instead are estimated based on the training data. For the purpose of re-ranking, we divide the items into two different groups based on their popularity, selecting the top 20\% in terms of interactions as the short-head or popular items, and the bottom 80\% as the long-tail or unpopular items. Furthermore, we compute a novelty score $N_{i}$ for each item, characterizing it as colder or warmer. In addition, we define the producer group fairness evaluation metric ($GF$) to capture the performance of models \textit{w.r.t.} the group-level fairness, and the sub-group cold-item fairness metric ($SGF$) to evaluate their performance on the sub-group level. $GF$ uses the deviation from producer group fairness $DPF$ presented in Equation \ref{eq:optimization} to compute the performance on the first level, while $SGF$ is a direct indicator of the average value of both sub-groups novelty scores. (cf. Section \ref{subsec:fairness-metrics})

We calculate the overall performance of our by averaging the accuracy, subgroup fairness (SGF), and group fairness (GF) metrics using the \squotes{All} metric, as shown in Table \ref{tbl:ext-results}, where $All = w_1 \cdot Acc + w_2 \cdot GF + (1-w_1-w_2) \cdot SGF$ where $w_1+w_2+w_3=1$. In this work, we set $w_i = \frac{1}{3}$, however, we can adjust the weights $w_i$'s to give more importance to specific evaluation metrics.

Furthermore, we employ the symbol $\Delta_B$ to indicate the percentage improvement between the \squotes{All}  metric of our modified \texttt{aPFR} models and the \textbf{Base} model, while $\Delta_t$ measures the percentage improvement between the \squotes{All} metric of our modified \texttt{aPFR} and the \textbf{\tPFR}.

\subsubsection{Core CF Recommendation Models}
We use a suite of competitive, CF recommendation models as baseline ranking models in our post-processing approach, as summarized below:

\begin{itemize}
    \item \textbf{BPR} \cite{rendle2012bpr}: A conventional recommendation model that employs matrix factorization to learn user and item embeddings of low dimensionality, and optimize the model based on the pairwise ranking of items for each user to predict whether a user prefers a given item over another;
    \item \textbf{NeuMF} \cite{he2017neural}: A hybrid recommendation model that combines matrix factorization with a neural network architecture (MLP). It learns user and item embeddings, capturing both linear and non-linear patterns in user-item interactions. The model uses a pairwise ranking objective to optimize the model;
    
    \item \textbf{MultiVAE} \cite{liang2018variational}: A non-linear probabilistic deep learning model that extends a variational autoencoder (VAE) structure to collaborative filtering for implicit feedback, and acquires the underlying representations of users and items from their interactions to create recommendations in an unsupervised way.

    \item \textbf{LightGCN} \cite{he2020lightgcn}: A pure collaborative filtering method that utilizes a simplified version of graph convolutional networks (GCNs) without nonlinear activation functions and additional weight matrices. It learns user and item embeddings through graph propagation rules and user-item interactions, making it scalable and efficient.

    \item \textbf{NGCF} \cite{wang2019neural}: A graph-based recommendation model that employs a neural network architecture and learns high-order connectivity and user-item signals based on the exploitation of the user-item graph structure, by propagating embeddings on it. 

\end{itemize}

\subsubsection{Hyperparameter Tuning.} The RecBole public library is utilized to implement and apply the baseline algorithms. Hyperparameter tuning is performed for both classical and deep recommendation models using a greedy search strategy. The best configurations are chosen based on the performance on the validation set. For BPR, we adjust the embedding size and learning rate hyperparameters, with 20 different trials, in the ranges [8, 16, 32, 64, 128] and [0.01, 0.005, 0.001, 0.0001], respectively. For NeuMF, we vary the learning rate, dropout probability, MLP hidden size, MF embedding size, and MLP embedding size hyperparameters, with 108 different trials, in the ranges [0.01, 0.005, 0.001], [0.1, 0.3], ['[64, 32, 16]', '[32, 16, 8]'], [64, 32, 16], and [64, 32, 16], respectively. For MultiVAE, we adjust the learning rate, latent dimension, MLP hidden size, and dropout probability hyperparameters, with 90 different trials, in the ranges [0.01, 0.005, 0.001], [8, 16, 32, 64, 128], [300, 600, 800], and [0.3, 0.5], respectively. For LightGCN, we vary the embedding size, learning rate, number of layers, and regularization weight hyperparameters, with 270 different trials, in the ranges [8, 16, 32, 64, 128], [0.01, 0.005, 0.001], [1, 2, 3, 4], and [1e-04, 1e-03, 1e-02], respectively. Lastly, for NGCF, we tune the learning rate, hidden size list, regularization weight, node dropout, message dropout, and delay hyperparameters, with 108 different trials, in the ranges [0.01, 0.005, 0.001], ['[64, 64, 64]', '[128, 128, 128]', '[256, 256, 256]'], [1e-5, 1e-4], [0.0, 0.1, 0.2], [0.0, 0.1, 0.2], and [1e-4, 1e-2, 1e-1], respectively.

\begin{table}[!h]
\centering
\caption{Harm values for different models and datasets at $\lambda = 0.1$, where $H \downarrow = 1 - N \uparrow$.}
\label{tbl:harm}
\begin{tabular}{cccc|cccc}
\toprule
& \multicolumn{3}{c}{Amazon Luxury Beauty} & \multicolumn{3}{c}{Foursquare}\\
\cmidrule{2-4} \cmidrule{5-7}
& BPR & NeuMF & LightGCN & BPR & NeuMF & NGCF \\
\midrule
Base & 0.23 & 0.24 & 0.22 & 0.27 & 0.28 & 0.3 \\
tPFR & 0.63 & 0.65 & 0.63 & 0.71 & 0.69 & 0.69 \\
LaPFR & 0.55 & 0.61 & 0.54 & 0.65 & 0.58 & 0.62 \\
MaPFR & 0.46 & 0.56 & 0.38 & 0.56 & 0.51 & 0.38 \\
HaPFR & 0.14 & 0.48 & 0.16 & 0.06 & 0.41 & 0.06 \\
\bottomrule
\end{tabular}
\end{table}
\begin{table*}[!h]
\centering
\caption{Detailed results for all models on the \texttt{Amazon Luxury Beauty} and \texttt{Foursquare} datasets at $\lambda = 0.1$.}
\label{tbl:ext-results}
\small
\scalebox{0.76}{\begin{tabular}{llllllllllllllllll}
\toprule
\multirow{2}{*}{Model} & \multirow{2}{*}{Type} & \multicolumn{7}{c}{Amazon Luxury Beauty} && \multicolumn{7}{c}{Foursquare} \\
\cmidrule{3-9}\cmidrule{11-17}
                        &                       & NDCG $\uparrow$ & GF$_{eq}$ $\uparrow$ & GF$_{prop}$ $\uparrow$ & SGF $\uparrow$ & All $\uparrow$ & $\Delta_B$ $\uparrow$ & $\Delta_t$ $\uparrow$ &&  NDCG $\uparrow$ & GF$_{eq}$ $\uparrow$ & GF$_{prop}$ $\uparrow$ & SGF $\uparrow$ & All $\uparrow$ & $\Delta_B$ $\uparrow$ & $\Delta_t$ $\uparrow$ \\
\midrule
\multicolumn{17}{c}{\textbf{}} \\ \hline
BPR & Base & 0.0092 & 0.5109 & 0.073 & 0.7679 & 0.5329 & - & - && 0.0071 & 0.5092 & 0.045 & 0.7283 & 0.5155 & - & - \\
BPR & tPFR & 0.008 & 0.8885 & 0.6878 & 0.369 & 0.5912 & 0.1094 & - && 0.0063 & 0.7308 & 0.2945 & 0.2952 & 0.4904 & -0.0487 & - \\
BPR & LaPFR & 0.0073 & 0.9701 & 0.8705 & 0.4495 & 0.6187 & 0.161 & 0.0465 && \cellcolor{purple!20}0.0061 & \cellcolor{purple!20}0.851 & 0.4938 & \cellcolor{purple!20}0.3466 & 0.5463 & 0.0597 & \cellcolor{purple!20}0.114 \\
BPR & MaPFR & 0.0066 & 0.9959 & 0.9618 & 0.5404 & 0.6314 & 0.1848 & 0.068 && 0.0055 & 0.9699 & 0.8534 & 0.4365 & 0.6001 & 0.1641 & 0.2237 \\
BPR & HaPFR & \cellcolor{purple!20}0.0067 & \cellcolor{purple!20}0.9967 & 0.9647 & 0.8627 & 0.7442 & \textbf{0.3965} & \cellcolor{purple!20}\textbf{0.2588} && 0.0018 & 0.998 & 0.9903 & 0.9359 & 0.6678 & \textbf{0.2954} & \textbf{0.3617} \\ \hline
NeuMF & Base & 0.0115 & 0.4791 & 0.024 & 0.7639 & 0.7425 & - & - && 0.0083 & 0.4965 & 0.0323 & 0.7216 & 0.5745 & - & - \\
NeuMF & tPFR & 0.0108 & 0.9857 & 0.9444 & 0.3551 & 0.7407 & -0.0024 & - && 0.0064 & 1.0 & 1.0 & 0.3131 & 0.5977 & 0.0404 & - \\
NeuMF & LaPFR & 0.0098 & 0.9857 & 0.9444 & 0.392 & 0.7129 & -0.0399 & -0.0375 && \cellcolor{purple!20}0.006 & \cellcolor{purple!20}1.0 & 1.0 & \cellcolor{purple!20}0.422 & 0.6201 & 0.0794 & \cellcolor{purple!20}0.0375 \\
NeuMF & MaPFR & \cellcolor{purple!20}0.01 & \cellcolor{purple!20}0.9857 & 0.9444 & \cellcolor{purple!20}0.4389 & 0.7379 & -0.0062 & \cellcolor{purple!20}-0.0038 && 0.0056 & 1.0 & 1.0 & 0.4855 & 0.6311 & \textbf{0.0985} & \textbf{0.0559} \\
NeuMF & HaPFR & 0.0082 & 0.9857 & 0.9444 & 0.5202 & 0.6898 & -0.071 & -0.0687 && 0.0042 & 1.0 & 1.0 & 0.5944 & 0.6257 & 0.0891 & 0.0468 \\ \hline
MultiVAE & Base & 0.0093 & 0.481 & 0.0 & 0.7426 & 0.5301 & - & - && 0.0111 & 0.4709 & 0.0061 & 0.7142 & 0.7226 & - & - \\
MultiVAE & tPFR & 0.0087 & 0.8203 & 0.44 & 0.2985 & 0.572 & 0.079 & - && 0.0097 & 0.7274 & 0.2923 & 0.269 & 0.5812 & -0.1957 & - \\
MultiVAE & LaPFR & \cellcolor{purple!20}0.0103 & \cellcolor{purple!20}0.9731 & 0.8198 & \cellcolor{purple!20}0.4076 & 0.7329 & \textbf{0.3826} & \cellcolor{purple!20}\textbf{0.2813} && \cellcolor{purple!20}0.0095 & 0.8838 & 0.5933 & \cellcolor{purple!20}0.3463 & 0.656 & -0.0922 & \cellcolor{purple!20}0.1287 \\
MultiVAE & MaPFR & 0.0074 & 0.9992 & 0.9693 & 0.5698 & 0.6761 & 0.2754 & 0.182 && 0.0056 & 0.9865 & 0.9478 & 0.5858 & 0.659 & -0.088 & 0.1339 \\
MultiVAE & HaPFR & 0.0052 & 0.9938 & 0.9316 & 0.8821 & 0.6878 & 0.2975 & 0.2024 && 0.0027 & 0.9994 & 0.9863 & 0.9657 & 0.704 & -0.0257 & \textbf{0.2113} \\ \hline
LightGCN & Base & 0.0092 & 0.545 & 0.0909 & 0.7761 & 0.5472 & - & - && 0.0074 & 0.5212 & 0.0592 & 0.7265 & 0.5359 & - & - \\
LightGCN & tPFR & 0.0095 & 0.835 & 0.5232 & 0.3719 & 0.6334 & 0.1575 & - && 0.0068 & 0.7165 & 0.2796 & 0.2817 & 0.4965 & -0.0735 & - \\
LightGCN & LaPFR & 0.0087 & 0.9686 & 0.8259 & 0.4601 & 0.6827 & 0.2476 & 0.0778 && \cellcolor{purple!20}0.0067 & 0.8416 & 0.4938 & \cellcolor{purple!20}0.3422 & 0.5581 & 0.0414 & \cellcolor{purple!20}0.1241 \\
LightGCN & MaPFR & \cellcolor{purple!20}0.0084 & \cellcolor{purple!20}0.995 & 0.9448 & \cellcolor{purple!20}0.6203 & 0.7342 & \textbf{0.3417} & \cellcolor{purple!20}\textbf{0.1591} && 0.0057 & 0.9617 & 0.8517 & 0.4925 & 0.6227 & 0.162 & 0.2542 \\
LightGCN & HaPFR & 0.0058 & 0.9754 & 0.8777 & 0.8401 & 0.6935 & 0.2674 & 0.0949 && 0.002 & 0.9961 & 0.9699 & 0.9433 & 0.6762 & \textbf{0.2618} & \textbf{0.3619} \\ \hline
NGCF & Base & 0.008 & 0.5117 & 0.0633 & 0.7694 & 0.4212 & - & - && 0.0052 & 0.5308 & 0.0871 & 0.7047 & 0.4075 & - & - \\
NGCF & tPFR & 0.0067 & 0.9187 & 0.7534 & 0.37 & 0.5478 & 0.3006 & - && 0.004 & 0.9769 & 0.8918 & 0.3104 & 0.5136 & 0.2604 & - \\
NGCF & LaPFR & \cellcolor{purple!20}0.0068 & \cellcolor{purple!20}0.9797 & 0.9131 & \cellcolor{purple!20}0.4642 & 0.6092 & 0.4463 & \cellcolor{purple!20}0.1121 && 0.0034 & 0.9834 & 0.9403 & 0.3782 & 0.521 & 0.2785 & 0.0144 \\
NGCF & MaPFR & 0.005 & 0.9902 & 0.9442 & 0.552 & 0.5672 & 0.3466 & 0.0354 && \cellcolor{purple!20}0.0045 & \cellcolor{purple!20}0.9862 & 0.9542 & \cellcolor{purple!20}0.6156 & 0.6364 & 0.5617 & \cellcolor{purple!20}0.2391 \\
NGCF & HaPFR & 0.0054 & 0.991 & 0.9465 & 0.8429 & 0.6798 & \textbf{0.614} & \textbf{0.241} && 0.0022 & 0.9865 & 0.9562 & 0.9442 & 0.6785 & \textbf{0.665} & \textbf{0.3211} \\
\bottomrule
\end{tabular}}
\end{table*}

\section{Findings}
\label{sec:findings}
This section details our experiments on several reproducibility aspects and presents our observations and results. 

\subsection{Sub-Group Fairness Harm Resulting from PFR}

Table \ref{tbl:ext-results} presents a summary of the primary outcomes from reproducing earlier research using \tPFR, as well as the proposed amendments (\laPFR, \maPFR, and \haPFR) across two datasets, namely \texttt{Amazon Luxury Beauty}, and \texttt{Foursquare}. Using this Table, we have created Table \ref{tbl:harm} in this section, which summarizes the negative impacts of each model (named Harm). To calculate the harm, we normalize sub-group novelty to a range of [0-1] across each dataset and compute it as $H=1-N$. Essentially, the concept is that the lower the sub-group fairness, the greater the harm to cold-items.

Based on Table \ref{tbl:harm}, it can be observed that adding \tPFR has a negative impact on sub-group novelty across all models, as indicated by higher H values compared to the base models. For example, in the case of the \texttt{Amazon Luxury Beauty} dataset, the H values increase from 0.23 to 0.63, 0.55, 0.46, and 0.14 for \tPFR, \laPFR, \maPFR, and \haPFR, respectively. This means that the proposed \haPFR model has the lowest harm value and introduces the least harm to sub-group novelty among the PFR variations. Similarly, for the \texttt{Foursquare} dataset, the H value for the base BPR model is 0.27, and this increases to 0.71, 0.65, and 0.56 for \tPFR, \laPFR, and \maPFR, respectively, indicating a negative impact on sub-group novelty. However, the \haPFR model has an H value of 0.06, much lower than the other PFR models, suggesting that it causes the least harm to sub-group novelty among the others.

\begin{figure*}[t]
    \centering
    \includegraphics[width=0.9\textwidth]{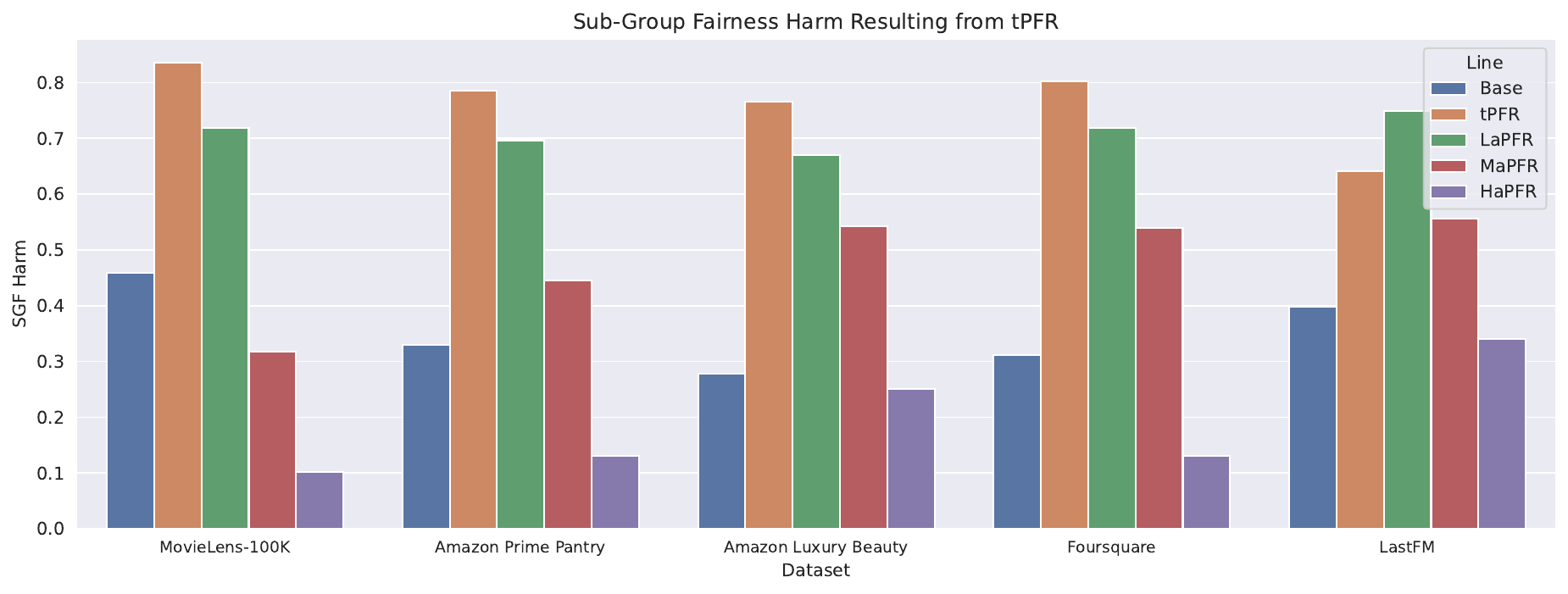}
    \caption{Sub-Group Harm resulting from traditional provider fairness re-ranking approach \tPFR and corrected via the proposed variations \laPFR, \maPFR, and \haPFR respectively. The figure shows the harm over five different datasets/domains and averages the results of five different baseline models.}
    \label{fig:harm}
\end{figure*}

Fig. \ref{fig:harm} presents a graphical representation of the harm inflicted by \tPFR on sub-group fairness harm across all five datasets. With the exception of the \texttt{LastFM} dataset, \tPFR is shown to have the highest harm H in nearly all cases. For the \texttt{LastFM} dataset, the lightest version of the amendment results in slightly higher harm to sub-group novelty, but this is remedied by \maPFR and \haPFR. This highlights the importance of balancing and regulating sub-group novelty and group-level fairness in producer fairness research.

\subsection{Proposed Amendment: \texttt{\textbf{aPFR}}}

We will first present the improvement in Table \ref{tbl:ext-results} with respect to the base ranker and the \tPFR, shown by $\Delta_B$ and $\Delta_t$, respectively.

\begin{itemize}

\item \textbf{Improvements with respect to base ranker ($\Delta_B$).} Positive changes can be noted in the $\Delta_B$-values, which represent the enhancement in the cumulative All metric compared to the Base (base ranker), for at least one of the versions of the proposed \texttt{aPFR}. For instance, on the Amazon dataset, BPR, MultiVAE, LightGCN and NGCF have the best values of 0.3965, 0.3826, 0.3417, and 0.614, respectively. Similar patterns can be seen in the Foursquare dataset, except for NeuMF (in Amazon) and MultiVAE (in Foursquare), which did not show an overall improvement across all three metrics with respect to the Base. Nevertheless, even in these cases, significant enhancements in recommendation fairness ($GF_{eq}$ and $GF_{prop}$) and $SGF$ were observed. For example, $GF_{eq}$ increased from 0.4791 to 0.9857 in Amazon with NeuMF, or $GF_{prop}$ increased from 0.0061 to 0.9863, while SGF increased from 0.7142 to 0.9657 in Foursquare with MultiVAE. These findings are intriguing because they indicate an average increase of $\Delta_B$ by 20 to 40\% compared to the base ranker, suggesting the effectiveness of our proposed amendment and its ability to achieve a better trade-off in comparison to the currently used fairness-unaware CF models. 


\item \textbf{Improvements with respect to the traditional PFR ($\Delta_t$).} The changes in $\Delta_t$-values, which measure the improvement of the cumulative All metric compared to \tPFR, are even more noteworthy. It can be seen that in several cases, at least one of the variants of the proposed \texttt{apFR} outperforms the others in terms of $\Delta_t$. Here are a few examples. On the Amazon dataset, BPR, MultiVAE, and NGCF have the best $\Delta_t$ values of 0.2588, 0.2813, and 0.241, respectively. Similarly,  in the Foursquare dataset, the values of $\Delta_t$ for BPR, LightGCN, and NGCF are 0.3617, 0.3619, and 0.3211, respectively. For the Amazon dataset, NeuMF does not show an overall improvement in this metric, while it offers only a slight improvement for the Foursquare dataset. Consistent with the trends observed in $\Delta_B$, in these cases, the GF and SGF metrics either maintain their values or show an improvement compared to \tPFR. For example, SGF increases from 0.3551 to 0.5202 in Amazon and from 0.3131 to 0.5944 in Foursquare, while $GF_{eq}$ and $GF_{prop}$ remain at 1 in Foursquare. In summary here again, the results show an improvement of $\Delta_t = 20-40\%$ compared to \tPFR in all experimental cases, which not only confirms the effectiveness of our proposed amendment to the classical \tPFR in solving the sub-group fairness problem but also demonstrates its ability to maintain or improve system accuracy and group fairness.
\end{itemize}

\begin{figure*}[t]
    \centering
    \includegraphics[width=\textwidth]{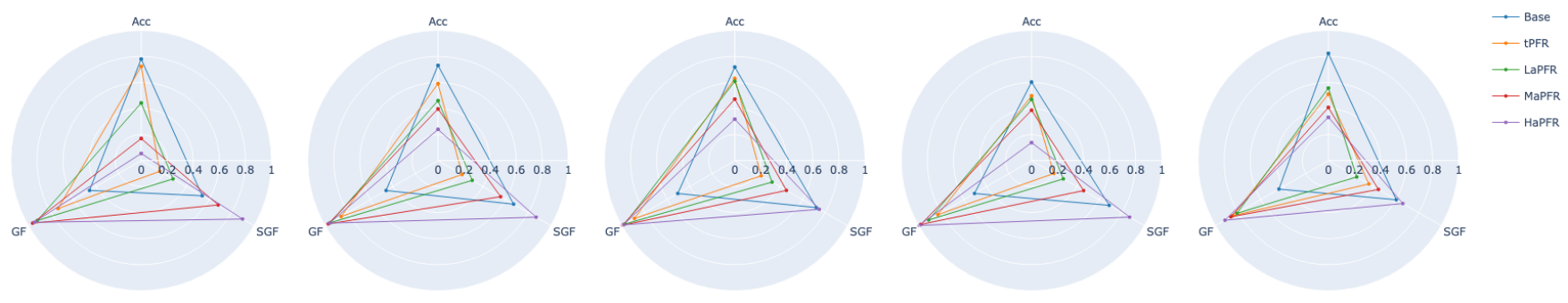}
    \caption{The Analysis of the Examined Accuracy, Group-Fairness, and Sub-group Fairness across the five datasets using the \textbf{average} of the five CF models in the five tested scenarios: base, \tPFR, \laPFR, \maPFR, and \haPFR.}
    \label{fig:tradeoff}
\end{figure*}

\begin{figure*}[t]
    \centering
    \includegraphics[width=\textwidth]{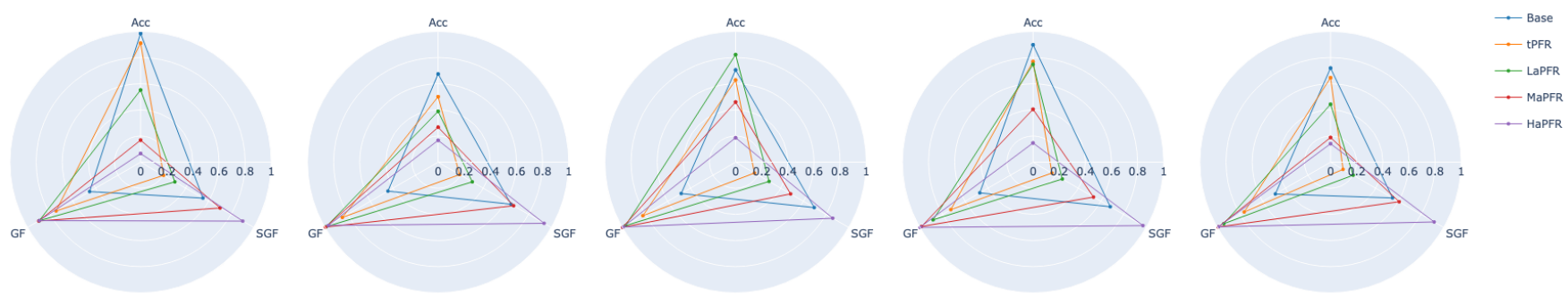}
    \caption{The Analysis of the Examined Accuracy, Group-Fairness, and Sub-group Fairness across the five datasets on the base \textbf{MultiVAE} model in the five tested scenarios: base, \tPFR, \laPFR, \maPFR, and \haPFR.}
    \label{fig:tradeoff-mvae}
\end{figure*}

\begin{figure*}[t]
    \centering
    \includegraphics[width=\textwidth]{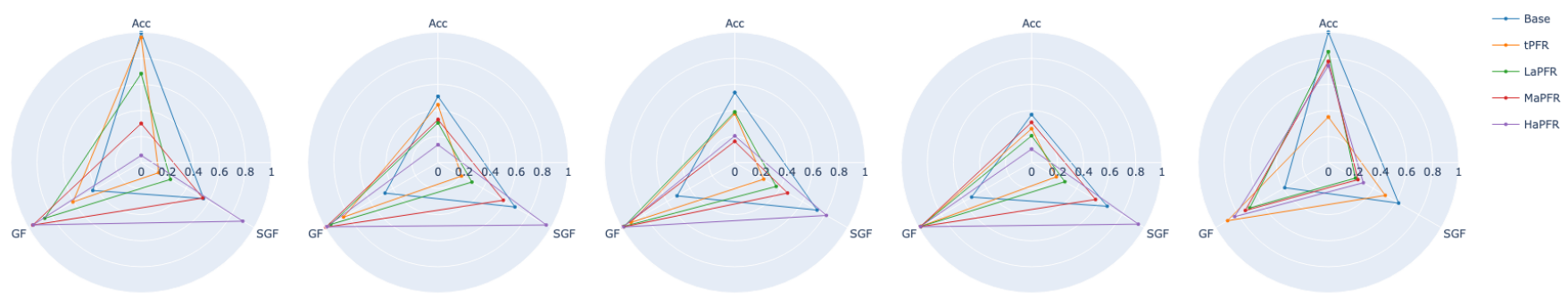}
    \caption{The Analysis of the Examined Accuracy, Group-Fairness, and Sub-group Fairness across the five datasets on the base \textbf{NGCF} model in the five tested scenarios: base, \tPFR, \laPFR, \maPFR, and \haPFR.}
    \label{fig:tradeoff-ngcf}
\end{figure*}

To gain an overall understanding of the relationship between the three evaluation objectives, we have created radar plots in Figures \ref{fig:tradeoff}, \ref{fig:tradeoff-mvae}, and \ref{fig:tradeoff-ngcf}, which illustrate the interplay between metrics as we transition from one dataset to another. In the radar plots, a larger triangle indicates a better model in terms of the underlying metrics. Figure \ref{fig:tradeoff} displays an overview of the amendment, which is calculated by taking the \textbf{average of the five base ranking models}. Figures \ref{fig:tradeoff-mvae}, and \ref{fig:tradeoff-ngcf}, on the other hand, focus on the methods MultiVAE and NGCF, respectively. It is evident that all versions of the proposed method are capable of improving SGF with respect to \tPFR, and even the base ranker for \haPFR or \maPFR. Moreover, being fair towards sub-groups increases group-level fairness as well. \laPFR, \maPFR, and \haPFR are observed to maximize GF almost in every case, surpassing the performance of \tPFR, which is a strategy specifically designed to increase GF. In Figure \ref{fig:tradeoff-mvae}, a more detailed analysis of the model MultiVAE is presented and it can be seen that SGF is better than the base ranker more frequently with \maPFR, while accuracy levels are maintained or even increased in some cases, such as Amazon Luxury Beauty with \laPFR. Similarly, Figure \ref{fig:tradeoff-ngcf} highlights the performance of NGCF, where it maintains high levels of accuracy with Foursquare, and \maPFR outperforms \tPFR in terms of SGF and GF. However, with LastFM, NGCF fails to raise SGF and GF compared to \tPFR, even though the traditional approach has the lowest accuracy. These results demonstrate the potential impact of datasets and specific data characteristics on the performance of a re-ranking method.

\subsection{Interplay between Re-Ranking Hyperparameters $\lambda$ and $\gamma$: an Ablation Study}

\begin{figure*}[!t]
    \centering
    \includegraphics[width=0.7\textwidth]{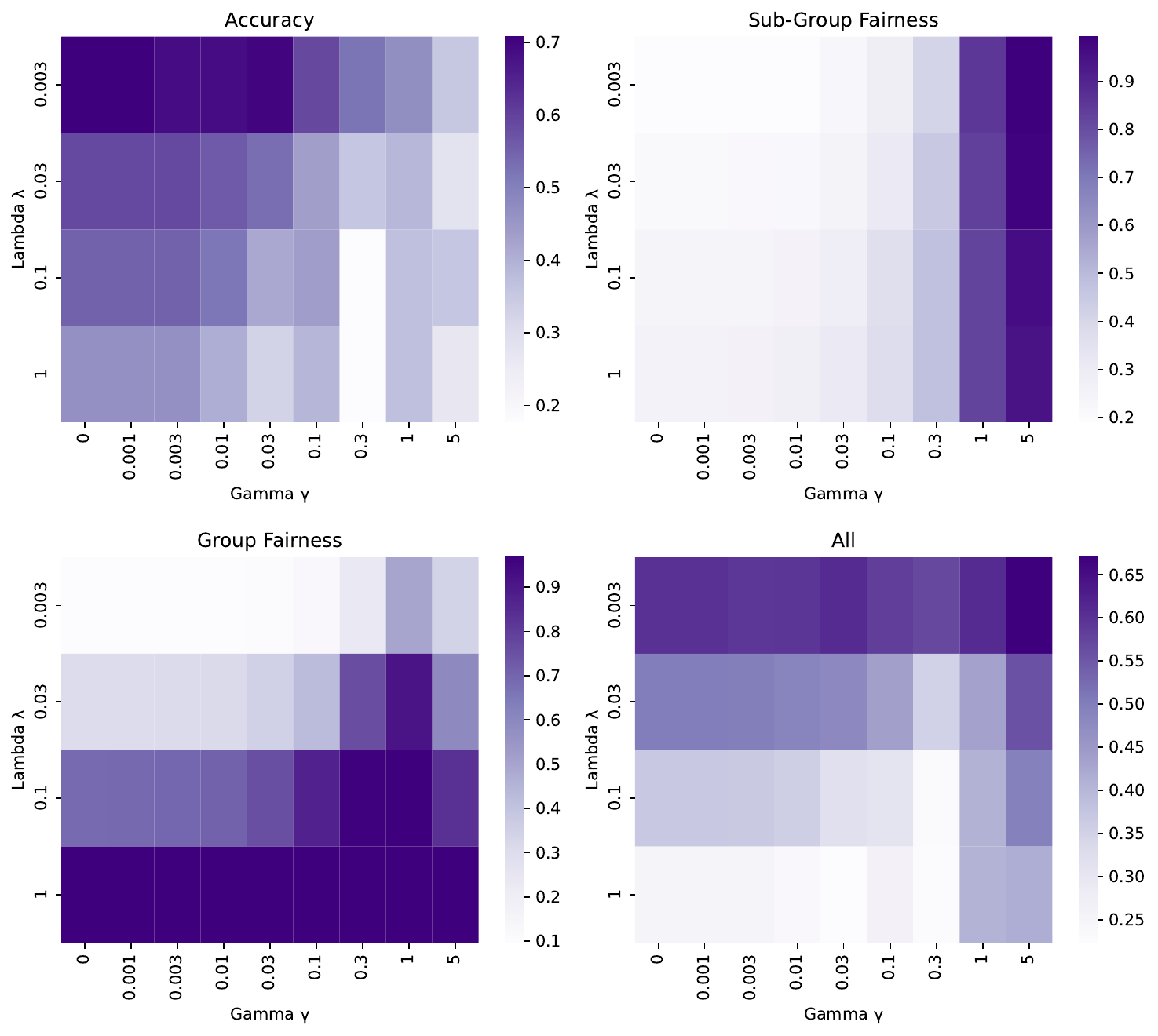}
    \caption{Interplay between the re-ranking hyperparameters $\lambda$ and $\gamma$ for the BPR model with the Amazon Luxury Beauty dataset.}
    \label{fig:interplay}
\end{figure*}

Figure \ref{fig:interplay} displays the relationship between the re-ranking hyperparameters $\lambda$ and $\gamma$ and their impact on accuracy, sub-group fairness, and group fairness. The first subplot indicates that reducing the values of both parameters increases system accuracy, as expected, due to the trade-off between optimization strategy and other dimensions. The second subplot reveals a linear relationship between sub-group fairness and novelty, while $\lambda$ has a negligible contribution. The third subplot demonstrates that higher values of $\lambda$ result in higher group fairness, but some higher $\gamma$ values can also increase both group and sub-group fairness. These results are consistent with those presented in Table \ref{tbl:ext-results}. Finally, the fourth subplot shows the average of the three metrics and dimensions.

\section{Conclusions and future work}
\label{sec:conc}

Our study focused on reproducing previous Producer Fairness Re-ranking (PFR) approaches with a spotlight on producer fairness. Delving deep into the works of \cite{naghiaei2022cpfair} and \cite{li2021user}, we identified how such methods could inadvertently harm colder items. This results in a fairness gap that spans across both advantaged and disadvantaged groups. Of relevance, recent works have taken \cite{li2021user} as a foundational reference, pushing the envelope further in areas such as consumer-producer fairness.

To counteract the challenges presented in sub-group fairness, we introduced a novel iteration of the conventional PFR method. This refined approach carefully balances accuracy and producer fairness, while prudently optimizing the selection of colder items within every group. Our experiments shed light on pivotal aspects, paving the way for future scholarly pursuits. Notably, the method we proposed accentuates sub-group fairness, enhances group-level fairness, and does so without compromising on accuracy. This underscores the pressing need to factor in individuals within groups when strategizing fairness-enhancing methods.

As we strive for more equitable recommender systems, it's crucial to look beyond the context we investigated. For instance, studies such as \cite{deldjoo2019assessing,anelli2021study} have assessed the impact of attacks on certain classes of users and items in recommender systems. Furthermore, each domain, e.g., the music industry \cite{deldjoo2023content,deldjoo2018content}, e-commerce, Point-of-Interest (POI), fashion \cite{deldjoo2023review}, multimedia, or even generative AI (\cite{deldjoo2023fairness}), might delineate harms to stakeholders in unique ways. Recent surveys and investigations, such as \cite{deldjoo2023content, rahmani2022unfairness, deldjoo2023fairness}, offer invaluable insights into these nuances. There is also increasing recognition of the need for a more unified methodology for fairness measurement in recommendation systems as proposed by works such as \citet{amigo2023unifying}.

Our research aligns with and advances the collective endeavor to render recommender systems that are both inclusive and equitable for all users. In the emerging landscape where fairness is paramount, our findings emphasize the importance of recalibrating approaches to ensure equity across the board.

\bibliographystyle{ACM-Reference-Format}
\bibliography{refs}

\end{document}